\documentclass{PoS}
\title{Diquark-antidiquark states with hidden or open charm}
\ShortTitle{Diquark-antidiquark states with hidden or open charm}
\author{Luciano Maiani\\
        Universit\`a di Roma `La Sapienza' and INFN 
        Sezione di Roma, Italy\\
        E-mail: \email{luciano.maiani@roma1.infn.it}}

\author{\speaker{Fulvio Piccinini}\\ INFN Sezione di Pavia and
        Dipartimento di Fisica Nucleare e Teorica\\ v. A. Bassi 6,
        I-27100 Pavia, Italy\\ E-mail:
        \email{fulvio.piccinini@pv.infn.it}}

\author{Antonio D. Polosa\\
         INFN Sezione di Roma and \\
        Universit\`a di Roma `La Sapienza', 
        Roma, Italy\\
        E-mail: \email{antonio.polosa@cern.ch}}

\author{Veronica Riquer\\
        INFN Sezione di Roma and \\
        Universit\`a di Roma `La Sapienza', 
        Roma, Italy\\
        E-mail: \email{veronica.riquer@cern.ch}}

\abstract{Some features and predictions of a recently proposed model based on diquark-antidiquark 
bound states are illustrated. Its ability in accomodating newly discovered 
charmed resonances around 4 GeV is discussed. 
}

\FullConference{International Europhysics Conference on High Energy Physics\\
		 July 21st - 27th 2005\\
		 Lisboa, Portugal}

\begin{document}

\section{Introduction}
The recent discovery~\cite{dsj} of the $j_q = 1/2$~P-wave charm-strange 
resonance $D_s^{**}$ with a mass significantly lower than expected 
has lead to new interest in charm spectroscopy. It was followed by the 
observation of some meson resonances with {\it no open} charm in $e^+ e^-$ 
annihilation with masses around 4 GeV starting with $X(3872)$~\cite{X3872} 
and so far ending with $Y(4260)$~\cite{Y4260}. These states can 
be fitted with difficulties in the charmonium $c {\bar c}$ picture (see 
ref.~\cite{quiggeps} for a recent review of alternative interpretations). 
On the other hand, the evidence of low mass resonances (below 1 GeV) 
by the KLOE Collaboration in the $\pi \pi$ spectrum of the radiative decay 
$\phi \to \pi_0 \pi_0 \gamma$, 
$\sigma(450)$~\cite{sigma}, and  by the E791 
Collaboration in the $K \pi \pi$ spectrum from 
$D$ decays, $\kappa(800)$~\cite{kappa}, 
reinforced the interpretation of the scalar nonet $(J^{PC}=0^{++})$ components below 
1 GeV, $f(980)$, $a(980)$, $\kappa(800)$ and $\sigma(450)$, as 4-quark bound 
states~\cite{jaffe}. In particular such states can be thought of as being $S-$wave 
bound states of a diquark and an antidiquark, $[q q] [\bar q \bar q]$, where the 
diquark is taken in the fully antisymmetric configuration ${\bf \bar 3_c}$, 
${\bf \bar 3_f}$ and ${\bf 1_s}$ of colour, spin and flavour respectively. 
The most convincing feature in favour of this interpretation is the inverted mass spectrum 
of the light scalar nonet, 
{\it i.e.} the lightest state has $I=0$ and no strangeness, 
while the heaviest particles have $I=1,0$ and like to decay in 
states containing strange quark pairs. In Ref.~\cite{mppr1} we have shown that this 
interpretation can reasonably describe the OZI allowed strong decays of the light 
scalar mesons in terms of a single amplitude parameterizing the switch of a $q \bar q$ pair. 

If the light scalar mesons are diquark-antidiquark composites, it is natural to consider 
analogous states with one or more heavy constituents~\cite{mppr2,mppr3}, {\it i.e.} of the 
form  $[c q] [\bar c \bar q']$, with $q, q' = u,d,s$. 
With respect to the light diquark case, two new elements come into the game: the 
near spin-independence of heavy quark forces (exact in the limit $m_c \to \infty$) 
and isospin breaking from light-quark 
masses. The first feature implies the presence of both spin zero and spin one diquarks, 
giving rise to a rich spectrum of states with $J=0,1,2$, with both natural and unnatural 
$J^{PC}$ quantum numbers. In the following we summarize the spectrum and decay properties 
of such four-quark states and their possible identification with recently discovered 
charmed resonances. 

\section{Spectrum of charmed diquark-antidiquark states}
\begin{figure}
\begin{center}
\includegraphics[width=0.35\textwidth]{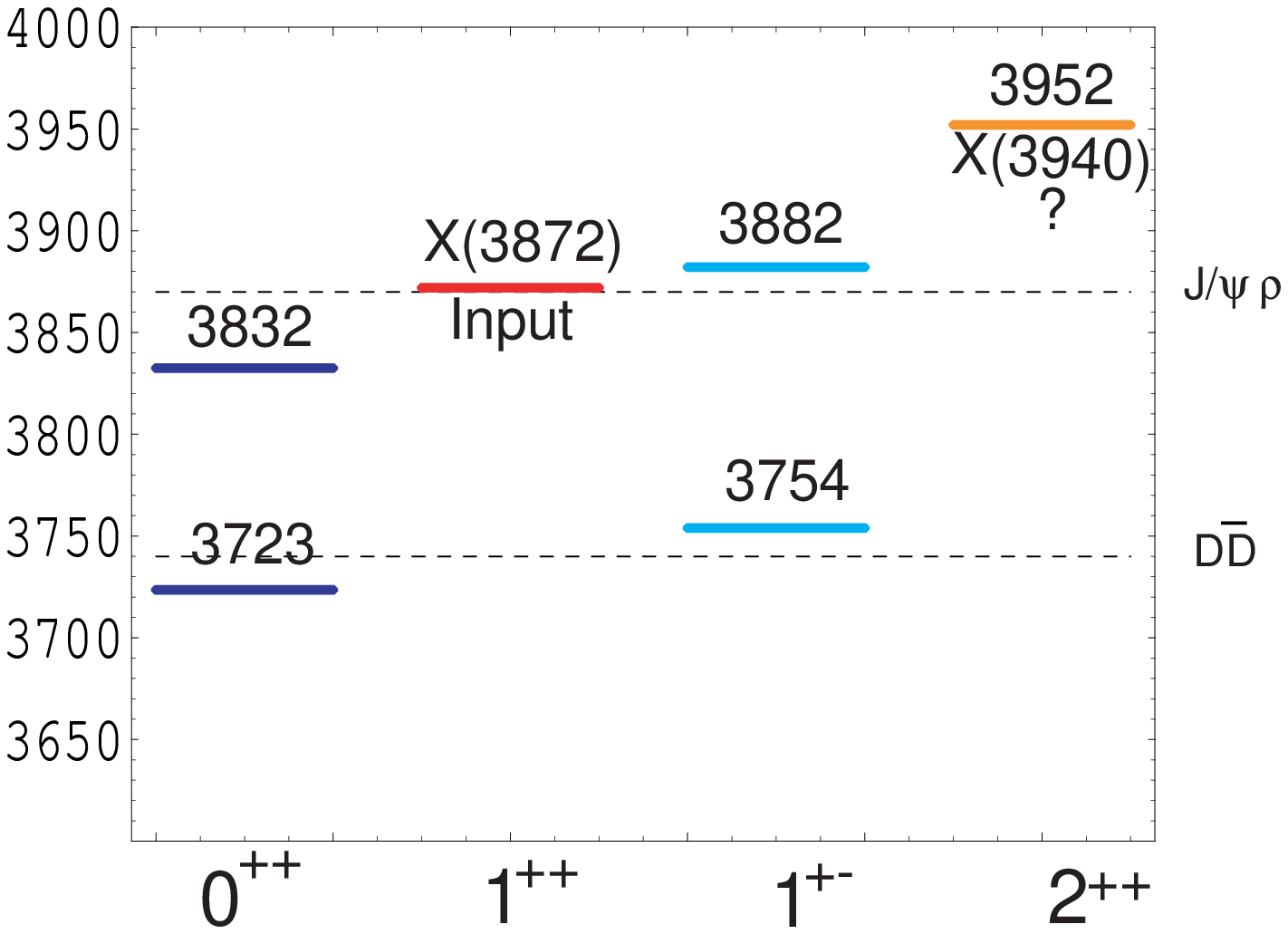}
\includegraphics[width=0.35\textwidth]{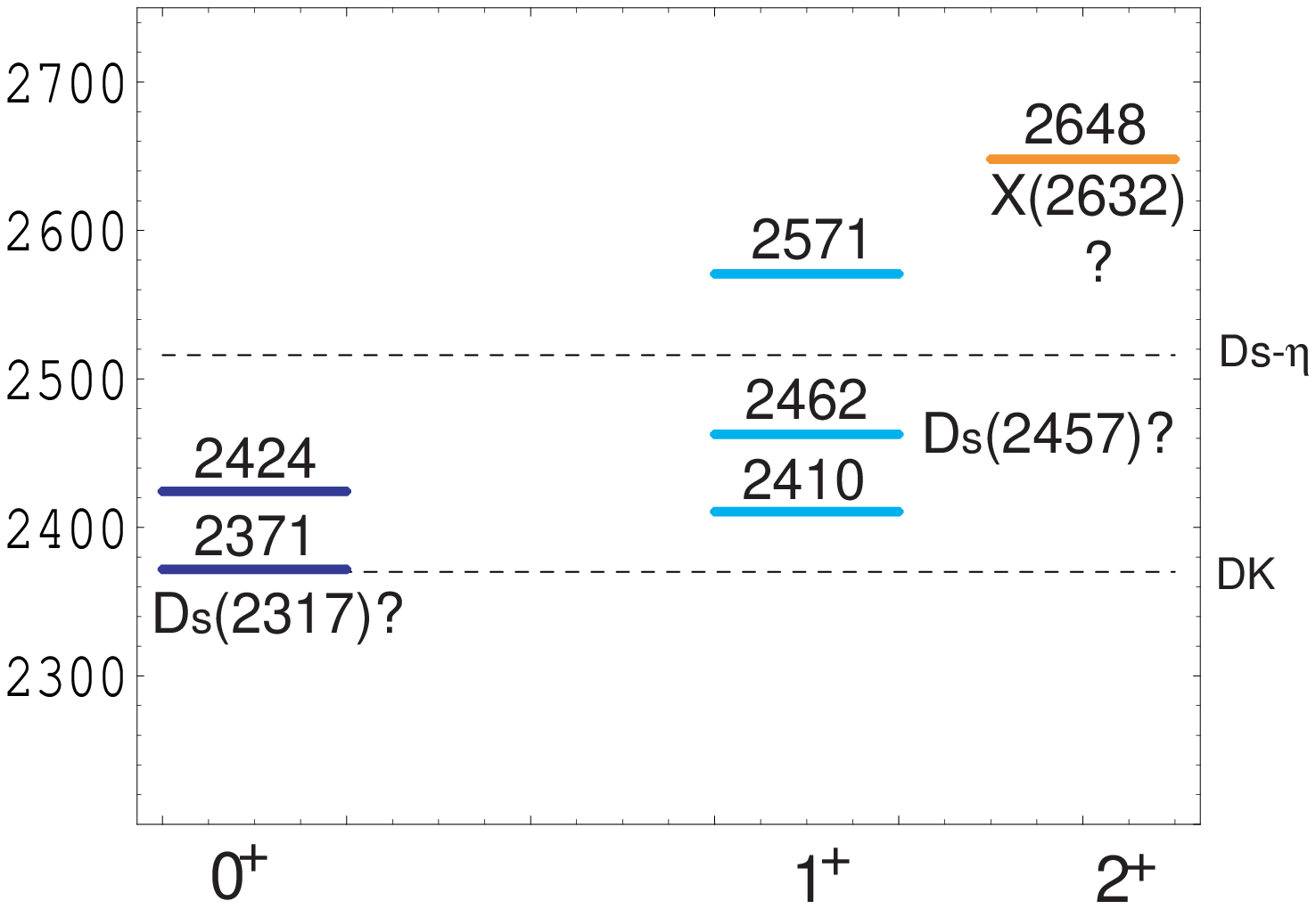}
\caption{The predicted spectrum of X particles with hidden charm (left) and with open 
charm and strangeness (right). Indicated mass values are in MeV. 
Dashed lines show the decay thresholds.}
\label{fig1}
\end{center}
\end{figure}

The mass spectrum of the systems $[c q] [\bar c \bar q']$ with $q, q'=u,d$ 
can be described  in terms of constituent diquark masses and spin-spin interactions, 
{\it i.e.} the Hamiltonian to be diagonalized is given by 
\begin{eqnarray}
H = 2 m_{[cq]} &+& 2 (\kappa_{cq})_{\bf \bar 3} [(S_c \cdot S_q) + (S_{\bar c}) \cdot 
(S_{{\bar q}'})] + 2 (\kappa_{q\bar q}) (S_q \cdot S_{{\bar q}'}) \nonumber \\
&+& 2 (\kappa_{c\bar q}) [(S_c \cdot S_{{\bar q}'}) + (S_{\bar c} \cdot S_q)] 
+ 2 (\kappa_{c\bar c}) (S_c \cdot S_{\bar c})
\label{hamilt}
\end{eqnarray}
and analogously for the $[c s] [\bar c \bar q']$ states. 
The Hamiltonian parameters can be obtained from known meson and baryon masses 
by resorting to the constituent quark model~\cite{cqm}
\begin{equation}
H = \sum_{i} m_i + \sum_{i<j} 2 \kappa_{ij} (S_i \cdot S_j), 
\label{cqm}
\end{equation}
where the sum runs over the hadron constituents. The coefficients $\kappa_{ij}$ depend 
on the flavour of the constituents $i, j$ and on the particular colour state of the 
pair. For instance, applied to the $L = 0$ mesons $K$ and $K^*$, Eq.~(\ref{cqm}) gives 
the relation $M = m_q + m_s + \kappa_{s\bar q} [J(J+1) - 3/2]$. Similar equations 
can be written for the meson pairs $\pi$-$\rho$, $D$-$D^*$ and $D_s$-$D_s^*$. 
Spin-spin interaction coefficients in antitriplet colour state can be estimated 
from baryon masses. For the quark-antiquark interactions to which we don't have yet 
experimental access we rely on estimates based on one-gluon exchange. 
The obtained results for the diquark masses are: $m_{[ud]} = 395$~MeV, 
$m_{[ud]} = 590$~MeV, $m_{[cq]} = 1933$~MeV. The explcit values for the spin-spin couplings 
$\kappa_{ij}$ are listed in Tab.~II and III of Ref.~\cite{mppr2}. By diagonalization of the 
Hamiltonian of Eq.~(\ref{hamilt}), on the basis of the states with definite diquark spin 
$S_{cq}$, antidiquark spin $S_{\bar c \bar q'}$ and total angular momentum $J$, we obtain 
six states corresponding to the following $J^{PC}$ assignments: $2 \times 0^{++}$, $1^{++}$, 
$2 \times 1^{+-}$, $2^{++}$. The state $1^{++}$, with the symmetric spin distribution 
$[cq]_{S=1} [\bar c \bar q]_{S=0}$+$[cq]_{S=0} [\bar c \bar q]_{S=1}$, 
is a good candidate to explain the properties 
of the $X(3872)$: it is expected to be narrow, like all diquark-antidiquark systems under 
baryon-antibaryon threshold; the unnatural spin-parity forbids the decay in $D$-$\bar D$, 
which is not seen; it can decay into both channels $J/\Psi \rho$ and $J/\Psi \omega$, 
as observed experimentally, due to isospin breaking in its wave function. 
By identification of the $1^{++}$ state with the observed $X(3872)$, the resulting 
mass spectrum is depicted in Fig.~\ref{fig1} (left). 
The analogous six eigenvectors of the Hamiltonian for 
the $[c q] [\bar s \bar q']$ are not invariant under $C$-conjugation and we have the 
following $J^P$ quantum numbers and multiplicities: $2 \times 0^+$, $3 \times 1^+$, $2^+$. 
The resulting spectrum is shown in Fig.~\ref{fig1} (right).
Of the two states $0^{++}$, one can decay in $\eta_c \pi$, $\eta_c \eta$ or multihadron 
states, the other one should decay into $D \bar D$. Up to now none of them has been 
seen experimentally. The same holds for the $1^{+-}$ states, which could decay in 
$J/\Psi + \pi (\eta)$, $\eta_c + \rho(\omega)$. The predicted mass of the $2^{++}$ state 
is, instead, in the right place of the $X(3940)$ seen by Belle in the channel 
$J/\Psi \omega$. In the discussed framework, however, it could also decay into 
$D^* \bar D^*$ and $D \bar D$ in $D$-wave. Of the states with strangeness, the lowest 
lying $0^+$ and one of the $1^+$ could be assigned to the $D_s(2317)$ and 
$D_s(2457)$~\cite{dsj}, respectively, 
in agreement with the observed decays $D_s(2317) \to D_s \pi^0$, 
$D_s(2457) \to D_s \gamma \pi^0$; $(D_s)^* \pi^0$. The state $2^+$ could fit naturally 
the particle claimed be the SELEX Collaboration~\cite{selex}.

\section{Isospin breaking}
Due to asymptotic freedom, at the large momentum scale implied by the heavy quark, 
the strength of the self-energy annihilation diagrams decreases. As a consequence, 
particle masses should be approximately diagonal with quark masses, even for up and 
down quarks~\cite{mppr3,rv}. In this limit the neutral mass eigenstate coincide with 
$X_u = [c u][\bar c \bar u]$ and $X_u = [c d][\bar c \bar d]$. Non-negligible 
gluon annihilation diagrams mix $X_u$ and $X_d$ giving rise to different eigenvalues 
separated by $\Delta M = 2 (m_d - m_u) / \cos(2 \theta) = (7\pm 2) / \cos(2 \theta)$ MeV. 
Thus the $X(3872)$ should consist, to a closer inspection, of two separate states. 
The mixing angle $\theta$ could be determined from $\Delta M$, as well from the 
ratio of the decay rates in $J/\Psi + \rho$ and $J/\Psi + \omega$, which are both 
allowed due to isospin breaking. From the measured ratio 
$\Gamma(J/\Psi \pi^+ \pi^-) / \Gamma(J/\Psi \pi^+ \pi^- \pi^0)$ by Belle we derive 
$\theta \simeq 20^\circ$, giving $\Delta M = 8 \pm 3$~MeV.

\section{Orbital excitations}
One of the distinctive features of the model based on diquark-antidiquark bound 
states is the presence of orbital angular momentum excitations, since 
the basic objects are coloured diquarks in a rising confining potential. 
Actually, the recently discovered state by BABAR $Y(4260)$~\cite{Y4260}, 
with $J^{PC} = 1^{--}$, could be one of such orbital 
excitations. It is seen in the channel $J/\Psi \pi^+ \pi^-$, with $M(\pi^+ \pi^-)$ around 
1 GeV, consistently with the decay $J/\Psi f_0(980)$. Given these features, the 
state can be interpreted as a bound state $[c s] [\bar c \bar s]$ with both diquarks 
in spin zero and with a unit of relative orbital angular momentum. In Ref.~\cite{mppr4} 
we have shown that this picture predicts a mass of $4330 \pm 70$~MeV, in nice agreement 
with the experimental value. A crucial prediction of the model is also the decay channel 
$D_s \bar D_s$, which still awaits for experimental observation.
\vskip 8pt
{\it Acknowledgements:} F.P. would like to thank the conveners for their invitation.

\end{document}